\documentstyle[11pt,newpasp,twoside,epsf]{article}
\def\edcomment#1{\iffalse\marginpar{\raggedright\sl#1\/}\else\relax\fi}
\reversemarginpar

\begin{document}

\title{The Physics of Blazar Optical Emission Regions I: 
Alignment of Optical Polarization and the VLBI Jet}

\author{Michael J. Yuan$^1$, Hien Tran$^2$, Beverley Wills$^1$, D. Wills$^1$}
\affil{$^1$University of Texas at Austin, $^2$Johns Hopkins University}

\begin{abstract}
We collected optical and near IR linear polarization data obtained over 20--30 
years for a sample of 51 blazars.  For each object, we calculated
the probability that the distribution of position angles was isotropic.  The
distribution of these probabilities was sharply peaked, with 27 blazars 
showing a probability $<$ 15\% of an isotropic distribution of position angles.  
For these 27 objects we defined a preferred position angle. 
For those 17 out of 27 blazars showing a well-defined radio structure angle
(jet position angle) on VLBI scales (1--3mas), 
we looked at the distribution of angle 
differences -- the optical polarization relative to the radio position angles.
This distribution is sharply peaked, especially for the BL Lac objects,
with alignment better than 15\deg\ for half the sample.
Those blazars with preferred optical position angles were much less likely to
have bent jets on 1--20mas scales.
These results support a shock-in-jet hypothesis for the 
jet optical emission regions. 
\end{abstract}

\section{Introduction}

Polarization observations have long been a very important probe of the
internal structure of blazar jets.  Bright spatially resolved knots often 
show radio polarization ({\bf E} vector) 
aligned with the projected jet direction,
indicating a perpendicular magnetic field.  This suggests that shocks are
responsible for compressing the jet magnetic field and accelerating the
synchrotron-emitting electrons (e.g. Aller, Aller \& Hughes 1985).

The relation between optical polarization and VLBI structure provides a
unique tool for investigating the regions of jet formation on $\la$
parsec scales. 
While previous statistical investigations have shown a tendency for optical
polarization to be aligned with the jet (Impey et al. 1991, 
Rusk \& Seaquist 1985), the 
interpretation of optical polarization is
less clear because the emitting regions in blazars are not resolved, blazars
often show violent short-term optical
polarization variability, and the old radio observations did not have 
sufficient angular resolution to probe the region near the optically emitting 
core.  
A few quasi-simultaneous optical-VLBI observations indicate that the
optical polarization
is aligned with the direction of newly-ejected blobs at the highest VLBI
resolutions (Gabuzda \& Sitko 1996, Lister \& Smith 2000).  
The optical polarization may originate in shocks at the
base of the jet.

We re-address the question of optical alignment,
taking advantage of a more extensive optical polarization database, and more
and improved VLBI maps.

\section{Data and Derived Parameters}

Our sample consists of 31 BL Lac type objects (BLLs) and 20 high polarization
QSOs (simply called QSOs) with extensive optical linear polarization data and
high quality VLBI maps taken from the literature.
Optical polarization data from observations spanning 20--30 years were
collected from the literature and McDonald Observatory archives.


We determined the following parameters for each blazar:

\begin{enumerate}
\item The probability that the measured optical polarization vectors are drawn
 from an isotropic distribution.
\item The preferred optical polarization position angle.  
This is the angle of the
vector average of the unit vectors corresponding to each 
polarization measurement.  We calculate this for objects with an 
isotropic distribution probability $<$ 15\%.
In these cases, our data are consistent with a single preferred angle (for an
exception with two preferred angles,
see the paper by Cross \& Wills, these Proceedings).
\item The position angle for the VLBI inner structure.   While some blazar jets
are straight, many are curved even on very small scales (Gomez et al. 1999,
Kellermann et al. 1998).   
Therefore, we measure position angles on both 1--3mas and
5--20mas scales, and determine a jet bending angle 
(the difference between them).
\end{enumerate}

\section{Results}

\begin{figure}[p]
\plottwo{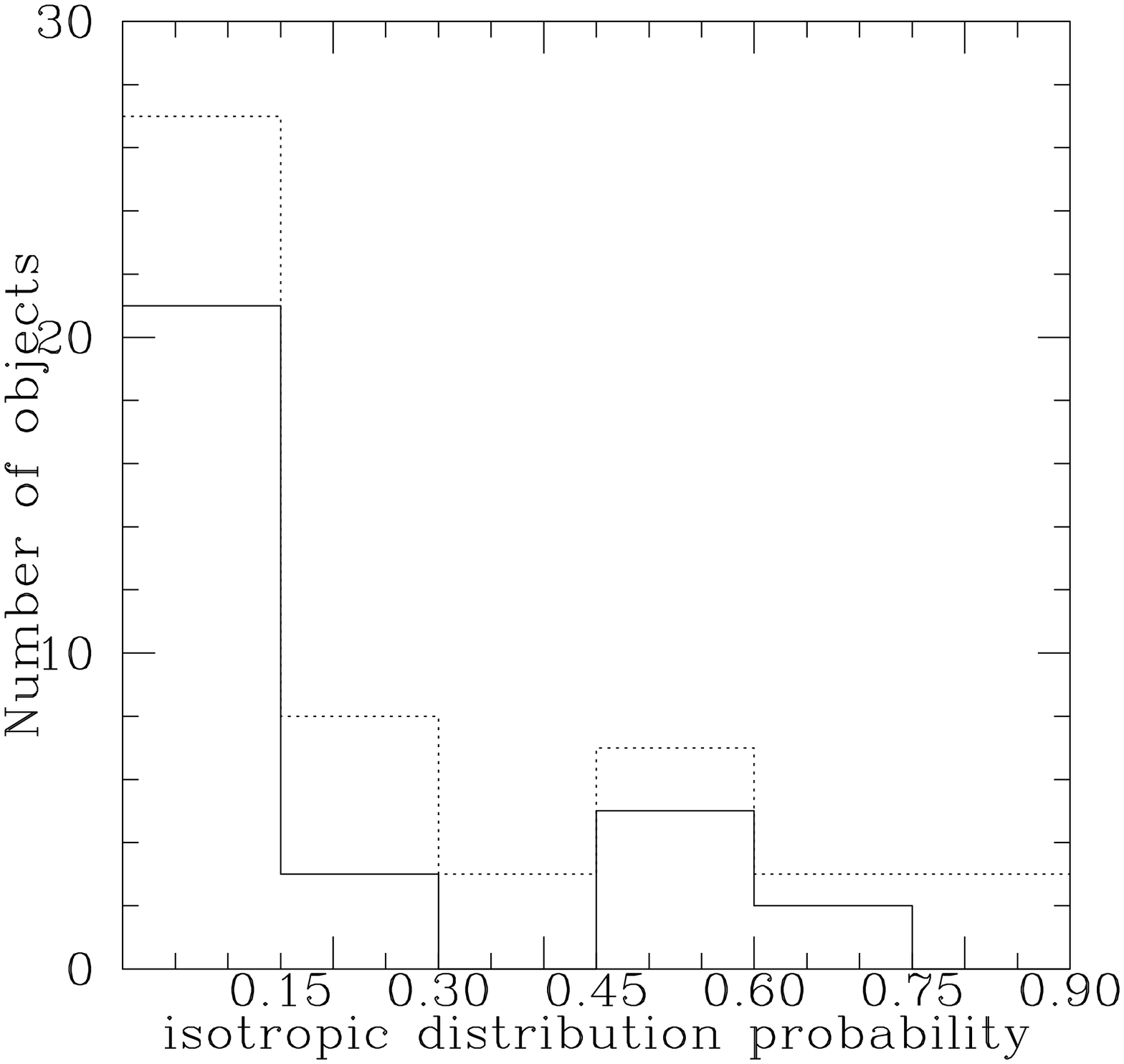}{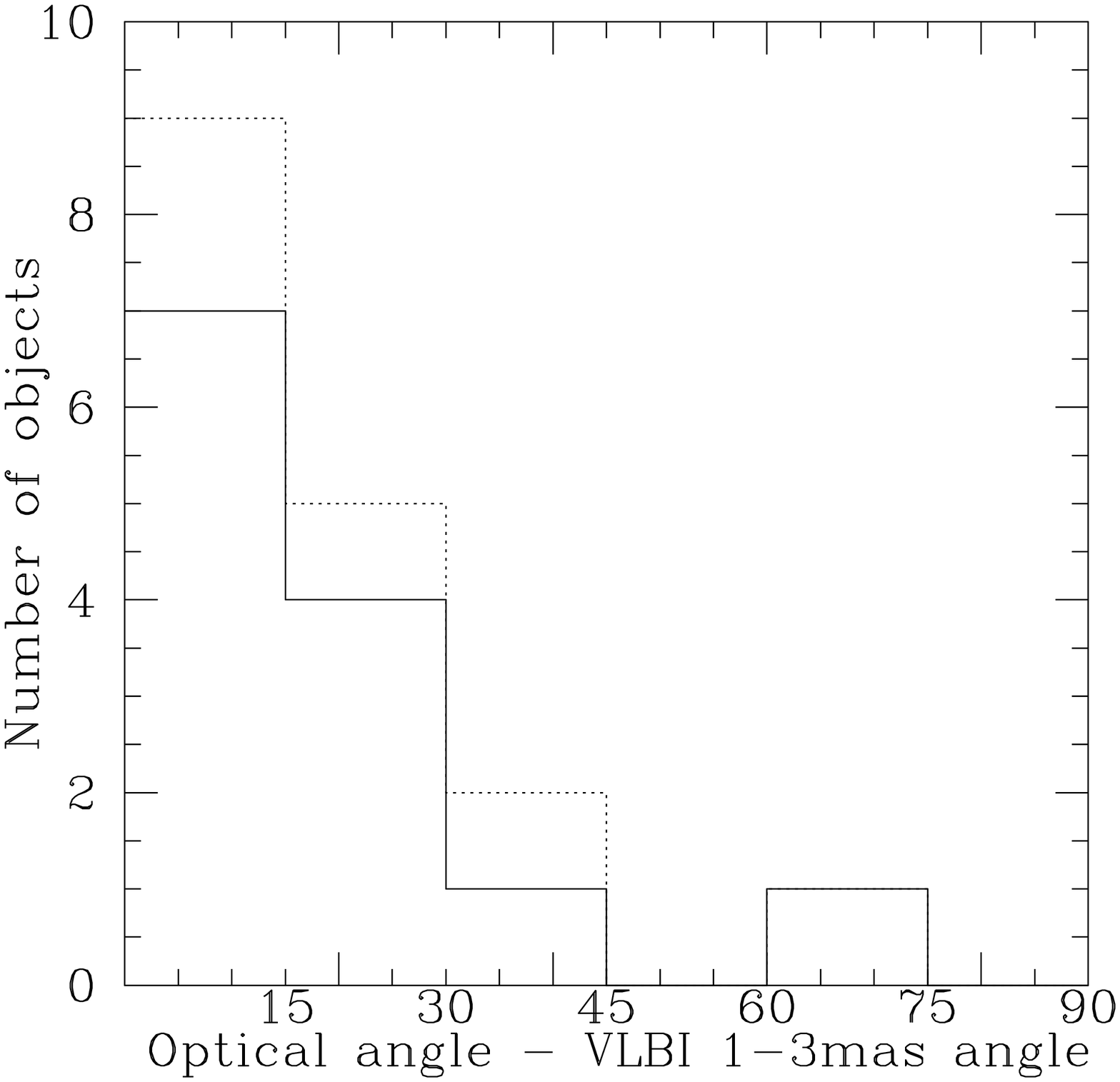}
\caption{Histograms of 
(left) isotropic distribution probability of
optical polarization position angles; (right) the difference between preferred
optical polarization and VLBI 1--3mas jet angle.
(Solid bars are for BL Lac objects and dashed bars are for BLLs and QSOs
combined). }
%
\vskip 10mm
\plottwo{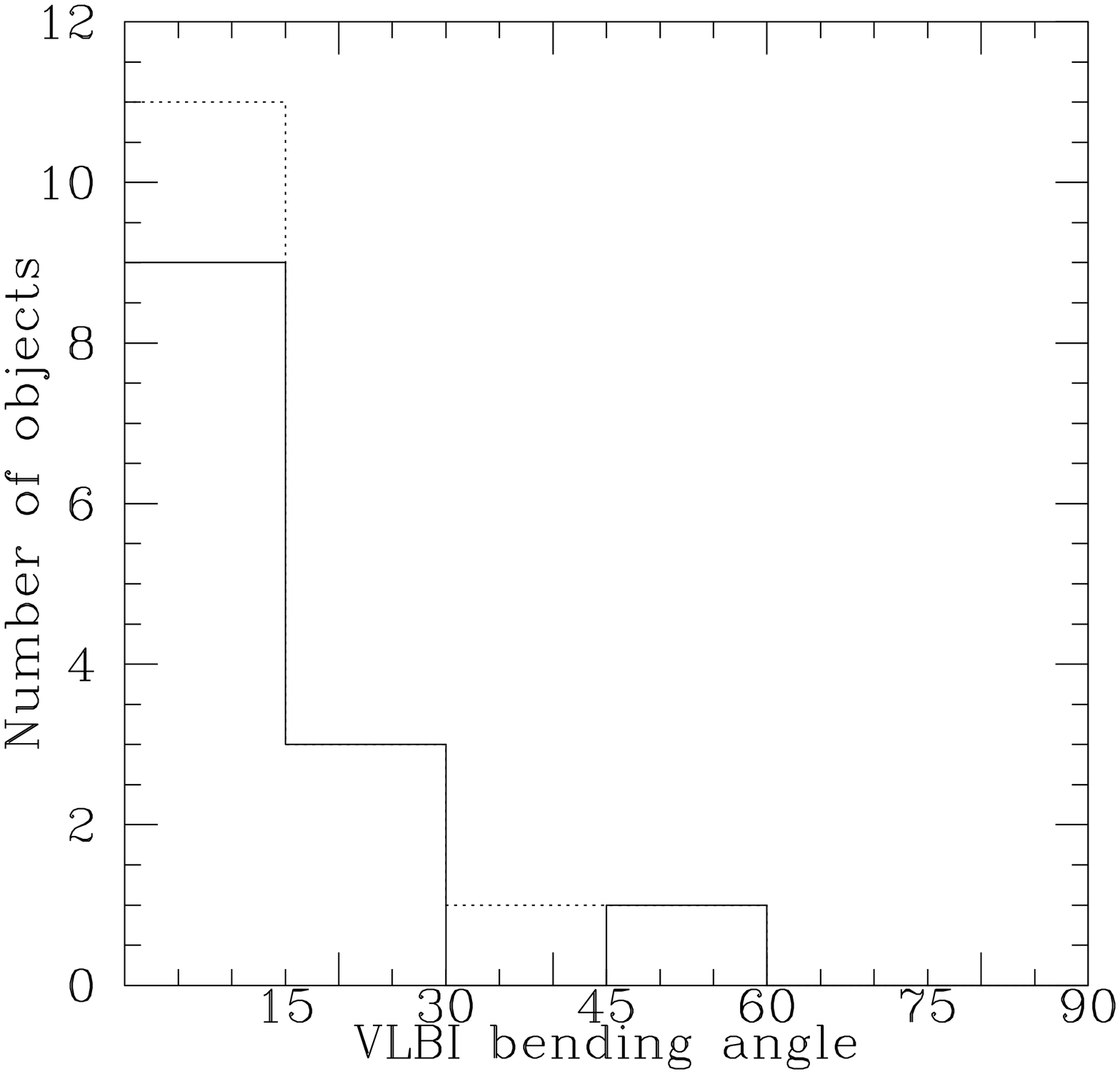}{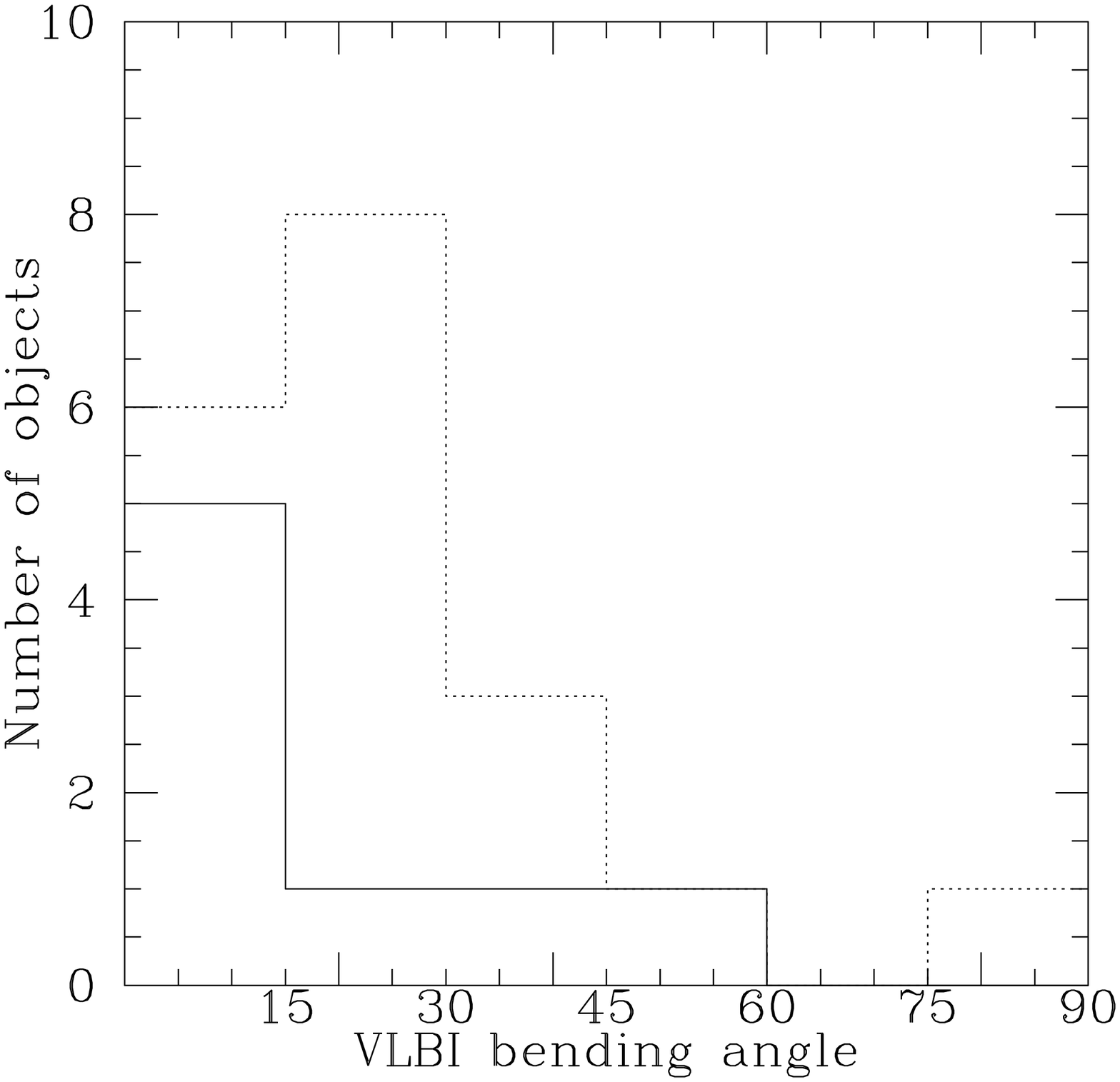}
\caption{Histograms of VLBI bending angle for (left) objects with
preferred optical polarization angles and for (right) objects without
preferred optical polarization angles.
(Solid bars are for BL Lac objects and dashed bars are for BLLs and QSOs
combined).  }
\end{figure}

\begin{enumerate}
\item Most BLLs show long-term, preferred optical-polarization angles 
despite their violent short-term variability (Figure~1--left).  The probability
of this distribution arising by chance is $<< 10^{-4}$ for BLLs alone, 
and for BLLs and QSOs combined.  The QSOs' distribution is significantly 
different from the BLLs' 
(0.5\% chance for the two to arise from the same underlying distribution) 
and consistent with an isotropic angle distribution.
\item When we look only at the objects with preferred optical polarization
position angles most BLLs have preferred optical polarization aligned
with the VLBI 1--3mas jet.  For BLLs, or BLLs and QSOs combined, 
the probability that Figure 1--right represents an isotropic distribution of
angles is $<$0.1\%.
\item The objects with preferred optical polarization angles show a 
strong tendency to have straight VLBI jets (bending angle
$< 15^{\circ}$) compared with objects with no preferred optical polarization
angles (Figure~2). The probability for the objects with preferred optical
angles to have the same VLBI bending angle distribution as the ones with no
preferred angles, is less than 1\%.
Objects with preferred optical polarization angle and small
VLBI bending are mostly BLLs. 


\end{enumerate}


\section{Discussion}

A natural explanation for the result that optical polarization
tends to align with the jet, is that the optical synchrotron emission
arises from a shock front in which the jet magnetic field has been compressed, 
on average, perpendicular to the jet.
The large scatter in the optical polarization angles for a given blazar suggests
that, in the inner jet region, the compressed magnetic field changes direction
with time. Possible explanations are that the inner 
jet is internally unstable, or
shocks may form via interaction with gas surrounding the 
central engine.  The jets
of QSOs may be affected by 
gas in the NLR and BLR, present in QSOs but absent in BLLs. This
may explain why QSOs
show preferred optical polarization angles less frequently.
The variations may be enhanced by the effects of foreshortening and relativistic
beaming.


The tendency that objects with  well-defined preferred optical polarization
directions also have very small VLBI scale jet bending
indicates that a well-behaved straight jet on parsec to Kpc scales corresponds
to a well-behaved jet on sub-parsec (optical) scales.
Large curvature is likely to be the effect of projection of small jet
curvature at very small viewing angles (Gower et al. 1982).  
Possible causes of jet
curvature are [1] an interaction with the environment, or [2]
an apparent curvature. In the first case, how does the base of the jet
know about the environment on much larger scales? 
The angular resolution of optical observations is
at best a factor of 100 worse than VLBI, often $> 100$mas. So we do not
have direct evidence to test our assumption that the more energetic 
optical photons arise near the base of the jet. The optical emission could
arise in the same shocks giving rise to cm-wavelength emission. 
The observation of rapid polarization variation at cm wavelengths, outside
the core, gives credence to this idea (Gabuzda et al. 2000).
In the second case, the direction of particle 
ejection may vary with time, for example, via a precession jet (e.g.
Hummel et al.
1997). Present data are inadequate to address changes in optical polarization
position angles on precession time scales.

\end{document}